\documentclass[
preprint,
amsmath,amssymb,
aps,
pra,
]{revtex4-1}
\usepackage{graphicx}
\usepackage{setspace}

\newcounter{constructcounter}

\newcommand{\fp}{fp}
\newcommand{\ket}[1]{\|#1\rangle}

\newenvironment{construct}[3]
{
	\begin{figure*}[tb]
	\begin{singlespace}
	\vspace{3mm}
	\setlength{\parindent}{0mm}
	\begin{minipage}{\linewidth}
	\refstepcounter{constructcounter}
	\rule{\linewidth}{0.3mm}
	\textbf{Construction \Alph{constructcounter}} #1
	
	\vspace{-3mm}
	\rule{\linewidth}{0.03mm}
	
	\vspace{1mm}
	\textbf{Input:} #2
	
	\textbf{Output:} #3
	
	\begin{enumerate}
}
{
	\end{enumerate}
	
	\vspace{-3mm}
	\rule{\linewidth}{0.03mm}
	\end{minipage}
	\vspace{3mm}
	\setlength{\parindent}{15pt}
	\end{singlespace}
	\end{figure*}
}

\begin{document}

\title{Analytical Error Analysis of Clifford Gates by the Fault-Path Tracer Method}

\author{Smitha Janardan}
\author{Yu Tomita}
\author{Mauricio Guti\'errez}
\author{Kenneth R. Brown}

\email{kenbrown@gatech.edu}

\affiliation{Schools of Chemistry and Biochemistry, Computational Science and Engineering, and Physics, Georgia Institute of Technology,
	Atlanta, GA 30332, USA}


\begin{abstract}
	
We estimate the success probability of quantum protocols composed of Clifford operations in the presence of Pauli errors.  Our method is derived from the fault-point formalism previously used to determine the success rate of low-distance error correction codes.  Here we apply it to a wider range of quantum protocols and identify circuit structures that allow for efficient calculation of the exact success probability and even the final distribution of output states. As examples, we apply our method to the Bernstein-Vazirani algorithm and the Steane [[7,1,3]] quantum error correction code and compare the results to Monte Carlo simulations.
	
\end{abstract}

\maketitle

\section{\label{sec:intro}Introduction}

As quantum information processors become more complex a key challenge is the validation and verification of integrated systems. Individual gates can be characterized by quantum process tomography \cite{Pauli_twirling_Chuang}, randomized benchmarking \cite{Emerson07,RB,RB3,RB4,Rand_Knill}, and related methods \cite{Flammia_tomo,Flammia_tomo2,Poulin_tomo,BlumeKohout2013,MerkelPRA2013,WallmanNJP2015}. Full characterization scales exponentially with the size of the gate , but efficient characterization is possible when the faulty gates have sparse descriptions \cite{qdyn_comp_sens} or only limited information, such as the average fidelity, is obtained \cite{RB5,RB6}. Efficient characterizations scale polynomially with the system size and can become impractical for larger gate sizes. An alternative method for testing larger devices is to compare the physical algorithmic output to the expected algorithmic output. For many algorithms, like the quantum linear system algorithm  \cite{HarrowPRL2009,CladerPRL2013}, the ideal output may not be known and the effect of errors on the output cannot be calculated.

Fortunately there are classes of quantum circuits that can be efficiently computed, with the prime example being circuits composed of only Clifford gates, which can be simulated efficiently by the Gottesman-Knill theorem \cite{CHP,Faster_than_CHP}. The circuits can then be decorated with random Pauli errors and the output can be sampled using Monte-Carlo methods. This Monte-Carlo sampling can be extended to include Clifford errors \cite{Cory} and Clifford gates conditional on measurements in a Pauli basis \cite{PRA_us,GutierrezPRA2015}. Since the Clifford group transforms Pauli errors to Pauli errors, all of the errors can be pushed to the end of the circuit. This transformation is the basis of fault-path methods which identify the sets of errors that result in failure by following how Pauli operators propagate through the correction circuit \cite{AGP_method}. For low-distance codes, these method are used to rigorously bound the fault-tolerant threshold of specific protocols. Exact calculations are not practical due to the exponential possible combinations of errors and these methods rely on cutoffs that consider only a certain number of errors. This is well motivated by the reduced probability of having multiple errors and the limited distance of the codes.  

Here we apply the fault-path method to algorithms made from Clifford circuits. While these algorithms provide at most only a polynomial advantage, they are ideal for testing the integration of many qubits into a quantum computer. Most quantum error correction codes expect that the errors are independent probabilistic Pauli operators. Implementing a non-fault tolerant circuit of Clifford gate and testing the output distribution relative to this model provides confidence in the accuracy of this error model for a given implementation.

In contrast to quantum error correction codes, we find that the fault-path method can efficiently calculate the exact success rate for certain tree-like quantum algorithms in polynomial time for Pauli error models. We show this can be determined from the graph structure of the circuit and discuss how the cost of exact simulation can be related to the weight of the nodes and the number of cycles in the graph. We then apply our tools to the Bernstein-Vazirani algorithm and exactly simulate the success rate for circuits containing up to 1350 qubits \cite{BValgo}. Finally, we apply error truncation to our method to estimate the threshold of the Steane [[7,1,3]] code with Shor ancilla \cite{Steane_QEC,TomitaPRA2013}.

\section{\label{sec:background} Background and Definitions: Pauli Errors and Clifford Circuits }  

\begin{figure}[tb]
	\begin{minipage}{0.65\textwidth}
		\includegraphics[width = 1\textwidth]{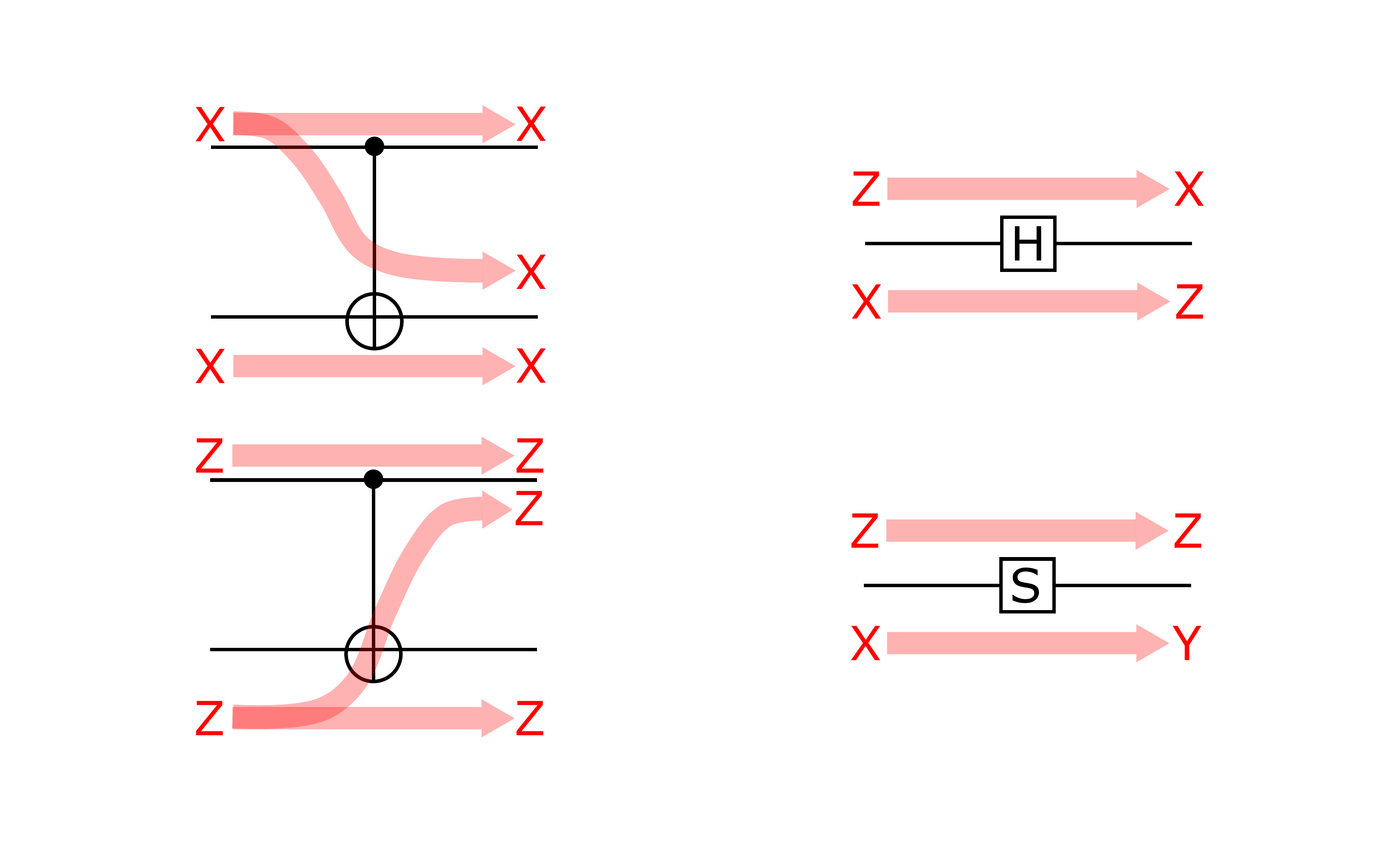}
	\end{minipage}
	\begin{minipage}{0.3\textwidth}
		\caption{\label{fig:error_prop} Two types of error ($X$ and $Z$) propagating across a controlled-NOT, Hadamard, and Phase gates.}
	\end{minipage}
\end{figure}

The Pauli operators on $n$ qubits are composed from the tensor product of the single qubit Pauli operators $X$, $Y$, and $Z$, and the Identity, $I$.  The weight of the Pauli operator is the number of non-identity elements in the tensor product. For $n$ qubits there are $4^n$ Pauli operators. The Clifford group is defined as unitary operations that transform Pauli operators to Pauli operators. The Clifford group can be generated from one and two qubit operations: \textit{CNOT}, $H$, and $S$. For additional information, we refer the readers to any quantum computation textbook \cite{MikeandIke}.

A Pauli error channel, $\mathbf{\mathcal{E}}$ is equivalent to a random application of a set of Pauli operators. The action of the channel is defined by Kraus operators $\mathbf{\mathcal{E}}(\rho)=\sum_j \mathbf{A}_j \rho \mathbf{A}^\dagger_j$, where $\mathbf{A}_j=\sqrt{p_j}\mathbf{P}_j$, $\mathbf{P}_j$ is a Pauli operator, and $p_j$ is the probability that the operator is applied. We define the number of non-zero $p_j$ as the rank of the channel, $r$. Clifford operators map Pauli error channels to Pauli error channels and although the weight of the Pauli operators can be changed the rank of the channel is preserved. Pauli error channels compose with other Pauli error channels to create new Pauli error channels with a rank that is bound by the product of the ranks of the channel or the maximum rank allowed by the system.

A standard model for errors is that each gate $g$ acting on $k$ qubits has an associated Pauli error channel $\mathbf{\mathcal{E}}_g$ composed of Pauli operators that also act on the same $k$ qubits, limiting the rank to $r_g \leq 4^k$. Assuming a circuit constructed from one and two-qubit Clifford operators, the maximum rank for each error channel is $16$. It is very convenient to push all of the error operators to the end of the circuit.  The other Clifford operations transform the error channel to $\mathbf{\mathcal{E}}^\prime_g$ but preserve the rank. If there are $G$ gates, the Pauli error channel of the entire circuit can be composed from $G$ Pauli error channels of low rank. The cost of this composition determines whether we can efficiently determine the probability distribution of outcomes and the success rate.

It is convenient to introduce the notion of an error vector, $\mathbf{\Psi}$, which contains the $4^k$ probabilities for a state to a specific Pauli error. Each Clifford gate, $g$, first transforms $\mathbf{\Psi}$ by mapping one Pauli error to another Pauli error. This can be represented by a $4^k \times 4^k$ transformation matrix $\mathbf{T}_g$, with only $4^k$ non-zero entries of 1 and preserving the error-free entry of the error vector. Then, the associated error channel $\mathbf{\mathcal{E}}_g$ is applied, which in this representation is a $4^k \times 4^k$ error matrix $\mathbf{E}_g$ which has $r_g$ distinct coefficients and $4^k r_g$ non-zero entries.  The transformation matrices for $H$, $S$, and \textit{CNOT} are given graphically in Fig. \ref{fig:error_prop}, alongside the full rank single qubit error matrix. To calculate the full error vector of $k$ qubits with $G$ gates, we can apply the formula: 

\begin{equation}\label{eq:transform_state}
	\mathbf{\Psi}_{final} = (\prod_{i=1}^{G} \mathbf{E}_i\mathbf{T}_i) \mathbf{\Psi}_{initial}.
\end{equation}
This calculation is impractical in general, but can be used for small problem sizes.

We often combine the error matrix and transformation matrix into a single bi-stochastic matrix: $\mathbf{M}_i = \mathbf{E}_i\mathbf{T}_i$.  As per Fig. \ref{fig:error_prop}, $H$ changes $X$ errors to $Z$ errors, Pauli operations such as $Z$ do not change Pauli errors.  Assuming the same error matrices for the two gates, we present two example bi-stochastic matrices:

\begin{equation*} 
	\mathbf{M}_{Z} =
	\begin{bmatrix}
		p_I & p_X & p_Y & p_Z \\
		p_X & p_I & p_Z & p_Y \\
		p_Y & p_Z & p_I & p_X \\
		p_Z & p_Y & p_X & p_I
	\end{bmatrix}, ~ \mathbf{M}_{H} =
	\begin{bmatrix}
		p_I & p_Z & p_Y & p_X \\
		p_X & p_Y & p_Z & p_I \\
		p_Y & p_X & p_I & p_Z \\
		p_Z & p_I & p_X & p_Y
	\end{bmatrix}
\end{equation*}

Let us examine two simple scenarios. In the first example, there are $G$ qubits each acted on by a single 1-qubit gate, and each gate has a distinct rank four error channel. In this case, every $\mathbf{\mathcal{E}}_g$ is equivalent to $\mathbf{\mathcal{E}}^\prime_g$, since there are no sequential Clifford gates. Finding the complete Pauli error channel requires multiplying all combinations of error probabilities to yield $4^G$ coefficients, which is inefficient in the circuit size. If we define the success probability as the probability of no qubits having error, we only need to consider the $I$ component of each error channel yielding a success rate, $P_{I,G}=\prod_g p_{I,g}$, which can be efficiently calculated with $G$ multiplications.

In a second example, there is one qubit with $G$ 1-qubit gates each with a distinct rank four error channel. Now the gates are in sequence and the channels are transformed by the gates to $\mathbf{\mathcal{E}}^\prime_g$.  Unlike the previous example, the final rank of the error channel is bound to be 4. We can compose two error channels by multiplying the 4 coefficients of each channel to yield only 4 coefficients. As a result the complete error distribution can be found efficiently with only $16 G$ multiplications of error probabilities after the error transformation. Generalizing to $k$ qubits, we require $16^k G$ multiplications, which is efficient in $G$ but inefficient in $k$. Formally, we calculate the bi-stochastic matrix for a sub circuit $F$.

\begin{equation} \label{eq:matrix_multi}
	\mathbf{M}_F = \prod_{g \in F} \mathbf{M}_g
\end{equation}

The crux of our method for calculating success rates is to cut every circuit into these two examples by identifying circuit components whose failure rate can be calculated independently and by limiting the size of the dependent block to a small numbers of qubits. If the circuit naturally has a small dependency, we can calculate the success rate exactly, otherwise we use approximations to truncate the dependency.

\section{Fault Path Method}

\begin{figure}[tb]
	(a) \includegraphics[width = 0.45\textwidth]{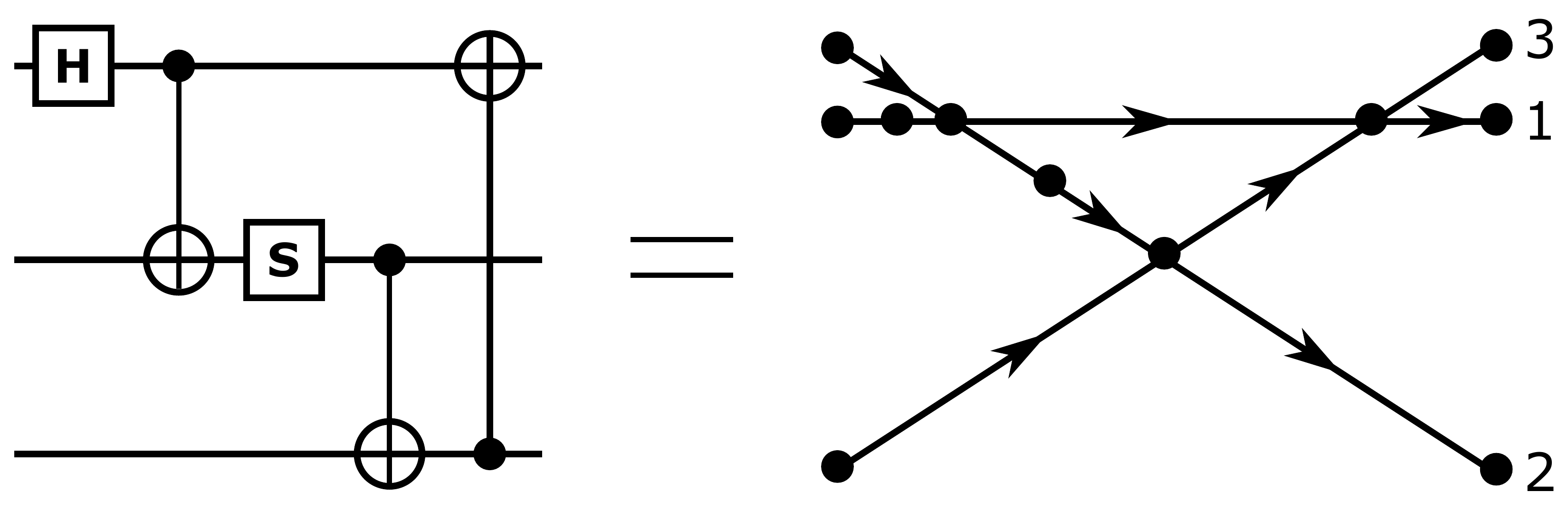} (b) \includegraphics[width = 0.45\textwidth]{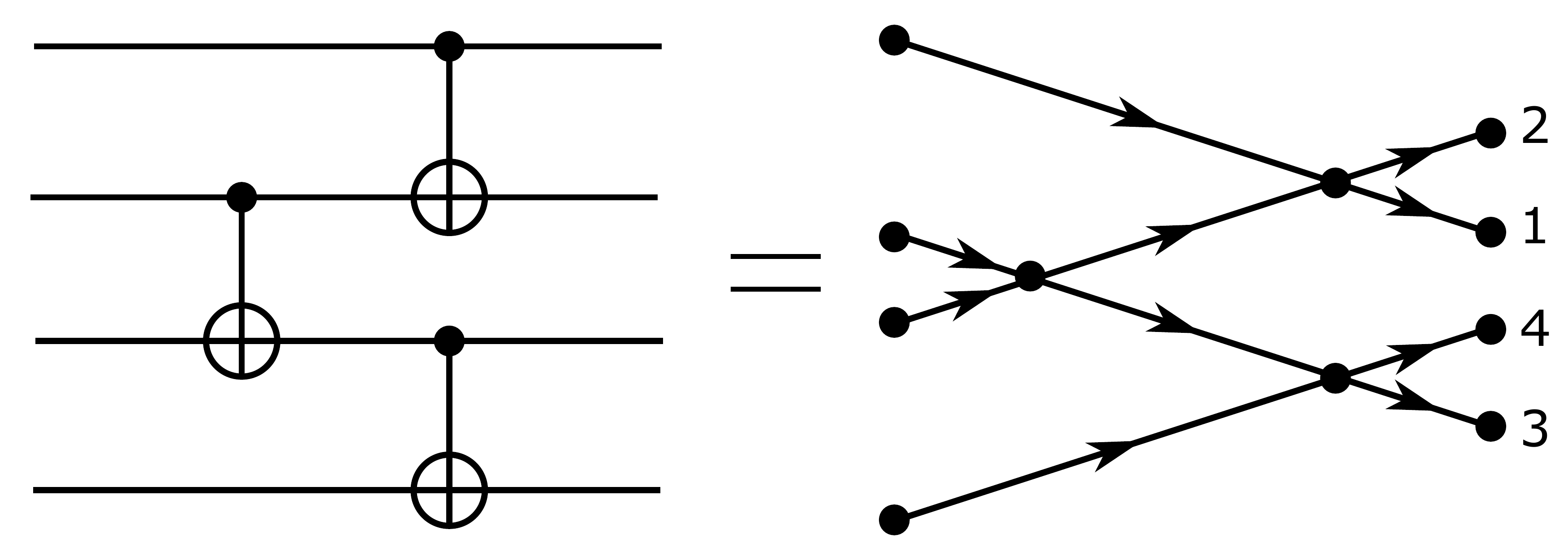}
	\caption{\label{fig:circuits} Demonstration of a standard circuit converted to a directed graph which contains: \textbf{a} a undirected cycle and \textbf{b} a tree-like pattern. Intersecting lines represent multi-qubit gates.}
\end{figure}

We start with a circuit of $G$ one and two-qubit gates. We convert the circuit to a directed graph where each gate is a node with incoming edges and outgoing edges corresponding to the qubits acted on by the gate. A fault path is defined by starting at an output qubit of the circuit and then walking the graph backwards to the input qubits. The fault-path shows where errors can arise that may propagate to the final qubit output.  We refer to our methods for using fault-paths to then calculate or estimate success rates as the Fault-Path Tracing (FPT) method.

Two circuits and their related graphs are shown in Fig. \ref{fig:circuits}. The fault-path, $\fp (q)$, finds all gates where errors can be introduced to the final state of qubit $q$ (Construction \ref{con:fault-path}). To calculate the error on that qubit for circuits composed of one and two-qubit gates, we break the fault-path into sub-paths of single qubit gates connected by two-qubit gates.  We can calculate the error matrix for the single-qubit gate paths efficiently as described earlier. Starting from the input nodes, we then combined these single qubit error matrices with the two-qubit error transformation matrix and gate error to generate a two-qubit error matrix on the outputs. We can then ask if the output qubit paths are in the fault path. If the answer is yes, we need to keep the two-qubit error matrix. If not, we can reduce the two-qubit error matrix to a one-qubit error matrix by tracing over the error state of the output qubit that is off the path. Either way, we then continue along the graph towards the output qubit. 

For tree-like graphs and a single fault path, we can always reduce to a single qubit error matrix after each gate. This simplification allows us to work with only single qubit error matrices except for at the two-qubit nodes where we need to calculate a two-qubit error matrix before reducing it. The result is an efficient method for calculating error states at single qubits without knowledge of the error states on other qubits (Construction \ref{con:qubit_success}). For undirected cycle on the underlying graph, the error matrices can continue to grow. In Fig. \ref{fig:circuits}a, we see that a two-qubit error matrix must be kept for a few nodes and that a three-qubit error matrix must be briefly constructed for the triangle-shaped loop. If we treat the undirected cycle as a single three-qubit Clifford gate, the graph becomes tree-like again but a three-qubit error matrix still must be generated. The number of qubits that input to the undirected cycle determines the size of error matrix that must be constructed.

For any algorithm, a lower-bound on the success probability can be determined by calculating the independent error probability of each output qubit having no error and then multiplying the probabilities.  This will overestimate the error since output errors on qubits will be correlated. In order to calculate the correlations, we need to look at the overlap between fault-paths that affect our output of interest. 

Our procedure for calculating error rates from overlapping fault paths is described in Construction \ref{con:gen_success}. The four cases mentioned are: error on no branch, error on control branch, error on target branch, and error on both branches. We often assume that the output qubit is measured in a specific basis $X$ or $Z$.  As a result, the fault path is simplified and reduces the Pauli errors to simply an error ($X$ or $Y$ for $Z$ measurements) or no error ($I$ and $Z$ for $Z$ measurements). We refer to this fault-path as $\fp (q;X) $. By breaking the overlapping fault-points into non-overlapping fault points, we can exactly calculate both the correlation and we can handle each subgraph exactly. However, in the case that there is a undirected cycle that has more than 2 qubit inputs or 2 qubit outputs, this method cannot no longer exactly calculate the success rate. Instead, a lower bound is used to estimate the rate for each subgraph.

\begin{construct}{\label{con:fault-path} Finding a fault-path}{The particular circuit being studied, $\mathbb{C}$; the qubit the fault-path is being formed from, $q$; and the assumed end error type, $\mathbb{E}$ if any.}{The fault-path containing a list of potential fault-points: $fp(q;\mathbb{E})$.}
	\item Find the last gate implemented on $q$ in $\mathbb{C}$ $\rightarrow$ $g$.
	\item The first fault-point is ($g$;$\mathbb{E}$). If no $\mathbb{E}$ specified, then two points ($X$ and $Z$).
	\item Based on $g$ and each $\mathbb{E}$, use the reverse error propagation rules to find all potential error sources $\rightarrow$ $\mathbb{S}$.
	\item \textbf{for source in $\mathbb{S}$ do}
	\item \hspace{10pt} Find the gate previous to $g$ that corresponds to the source, which may or may not be on the same $q$ $\rightarrow$ $g$.
	\item \hspace{10pt} Determine the new error type after error transformation $\rightarrow$ $\mathbb{E}$.
	\item \hspace{10pt} $fp(q;\mathbb{E})$ += ($g$;$\mathbb{E}$).
	\item \textbf{end for}
	\item Repeat steps 3-8 until reached the beginning. 
\end{construct}

\begin{construct}{\label{con:qubit_success} Probability of success for single tree-like fault-path}{The fault-path for a single qubit, $fp$; a dictionary of bistochastic matrices for each gate on the fault-path, $\mathbb{M}$.}{Probability of the qubit yielding the correct output, $\bar{\varepsilon}$.}
	\item Ensure that points in $fp$ are well-ordered based on the circuit, $\mathbb{C}$.
	\item Let $\Psi$ be [1,0,0,0].
	\item \textbf{for gate in $fp$ do}
	\item \hspace{10pt} Find the matrix in $\mathbb{M}$ corresponding to the gate.
	\item \hspace{10pt} If the gate is a two-qubit gate, condense the matrix to a 4x4 matrix.
	\item \hspace{10pt} Apply the matrix to $\Psi$ using \textbf{Eq. \ref{eq:matrix_multi}}.
	\item \textbf{end for}
	\item $\bar{\varepsilon}$ is the first element in $\Psi$.
\end{construct}

\begin{construct}{\label{con:gen_success}Approximate probability of success for multiple fault-paths}{The list of fault-paths, $\mathbb{F}$.}{Probability of all fault-paths having no error, $\bar{\varepsilon}$, and the probability of all fault-paths having error, $\varepsilon$.}
	\item Split $\mathbb{F}$ into independent groups.
	\item \textbf{for each independent group do}
	\item \hspace{10pt} Find the fault-points common to all fault-paths, and remove points from each path.
	\item \hspace{10pt} Determine if all fault-paths (with common points removed) separate into $n$ independent branches.
	\item \hspace{10pt} For the common fault-points, find the probabilities of all $2^n$ possible cases (all combinations of each branch having or not having error).
	\item \hspace{10pt} Build the probability state, $\mathbf{\Psi}$, from these rates.
	\item \hspace{10pt} \textbf{IF} Branches are independent \textbf{THEN} call \textbf{Construction \ref{con:gen_success}} for each branch.
	\item \hspace{10pt} \textbf{IF} Branches are dependent (therefore part of a cycle) \textbf{THEN} use $\varepsilon = \prod_{i} \epsilon_i$, for each fault-path (worst case).
	\item \hspace{10pt} Using the branch error rates, build a stochastic matrix.
	\item \hspace{10pt} Apply matrix to $\mathbf{\Psi}$.
	\item \hspace{10pt} $\mathbf{\Psi}$[first] $\rightarrow$ Success rate.
	\item \hspace{10pt} $\mathbf{\Psi}$[last] $\rightarrow$ Error rate.
	\item \textbf{end for}
	\item $\bar{\varepsilon}$ is the product of all independent groups success rates.
	\item $\varepsilon$ is the product of all independent groups error rates.
\end{construct}

\subsection{\label{sec:BV}Bernstein-Vazirani Algorithm}

\begin{figure}[tb]
	\begin{minipage}{0.5\textwidth}
		\includegraphics[width = \textwidth]{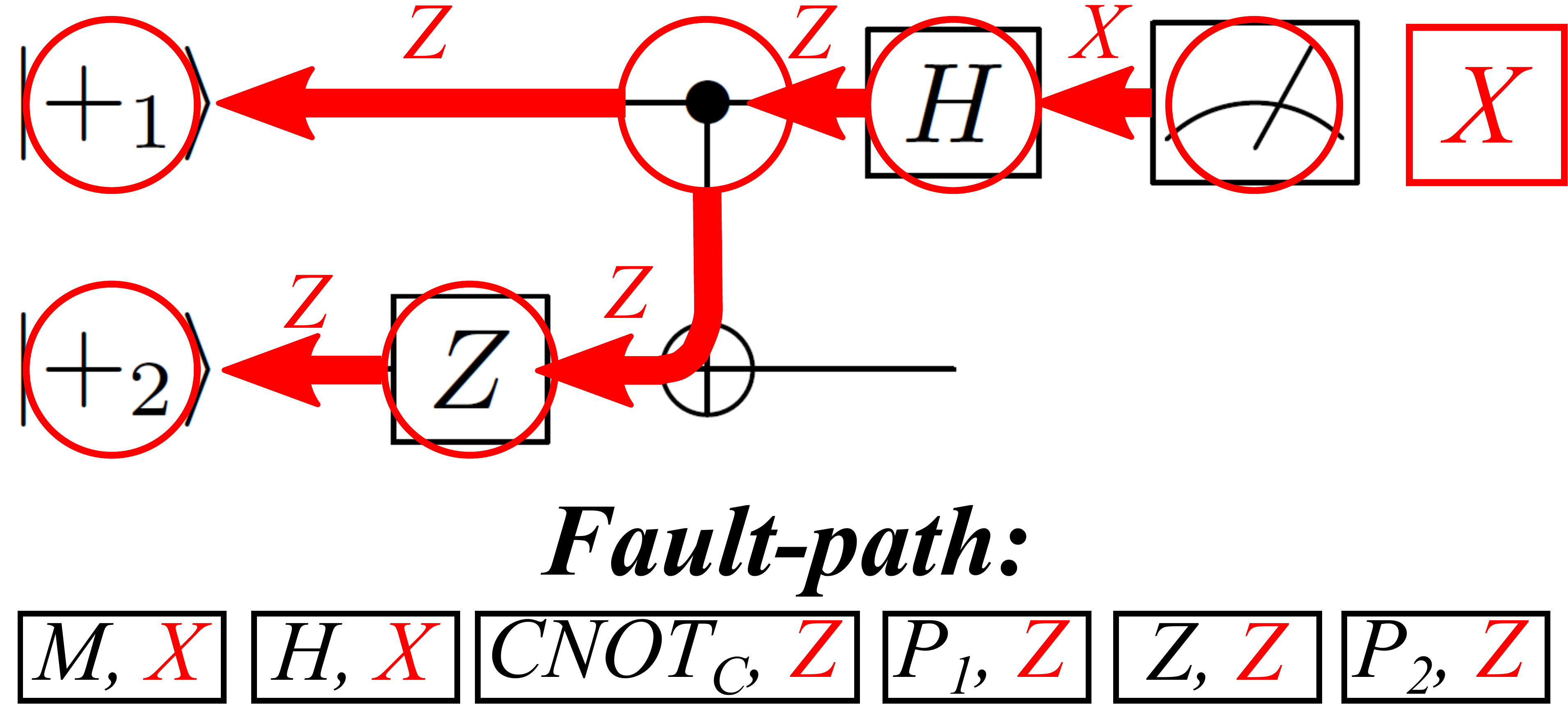}
	\end{minipage}
	\begin{minipage}{0.05\textwidth}
	\end{minipage}
	\begin{minipage}{0.45\textwidth}
		\caption{\label{fig:faultpath_example} The Bernstein-Vazirani Algorithm for one bit, with a Hamming weight of one. A possible X error on the first qubit could have resulted from various previous gates, found through backwards error propagation rules. The controlled-X gate leads to a branch in the fault-path. Each possible error source is a fault-point and has an associated error type that affects the output.}
	\end{minipage}
\end{figure}

\begin{figure}[tb]
	(a) \includegraphics[width = 0.45\textwidth]{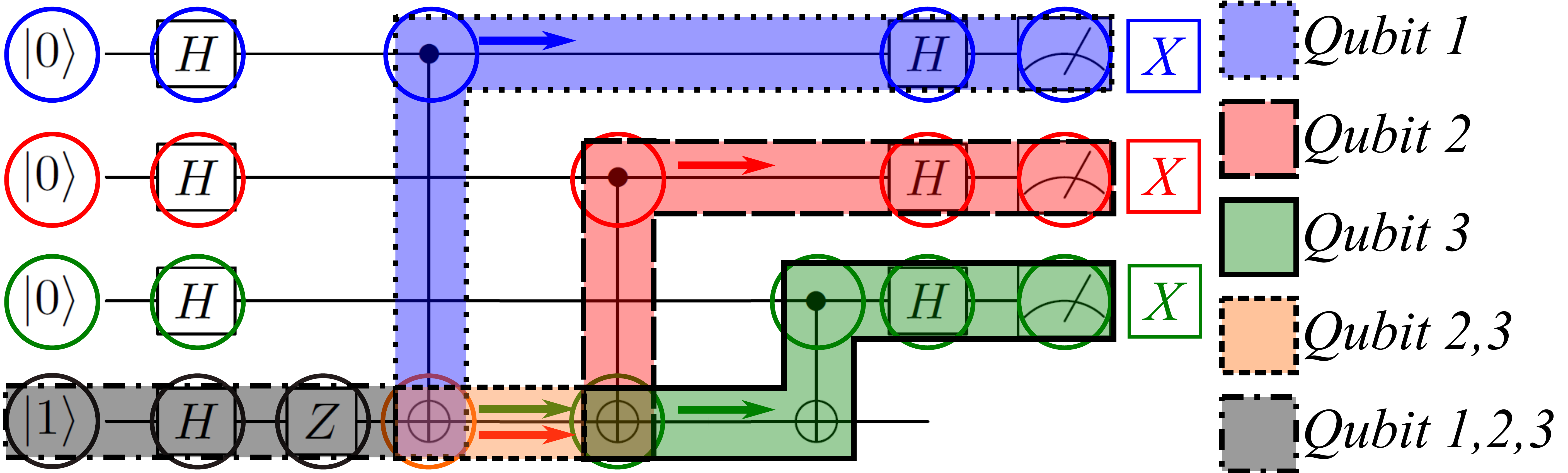} (b) \includegraphics[width = 0.45\textwidth]{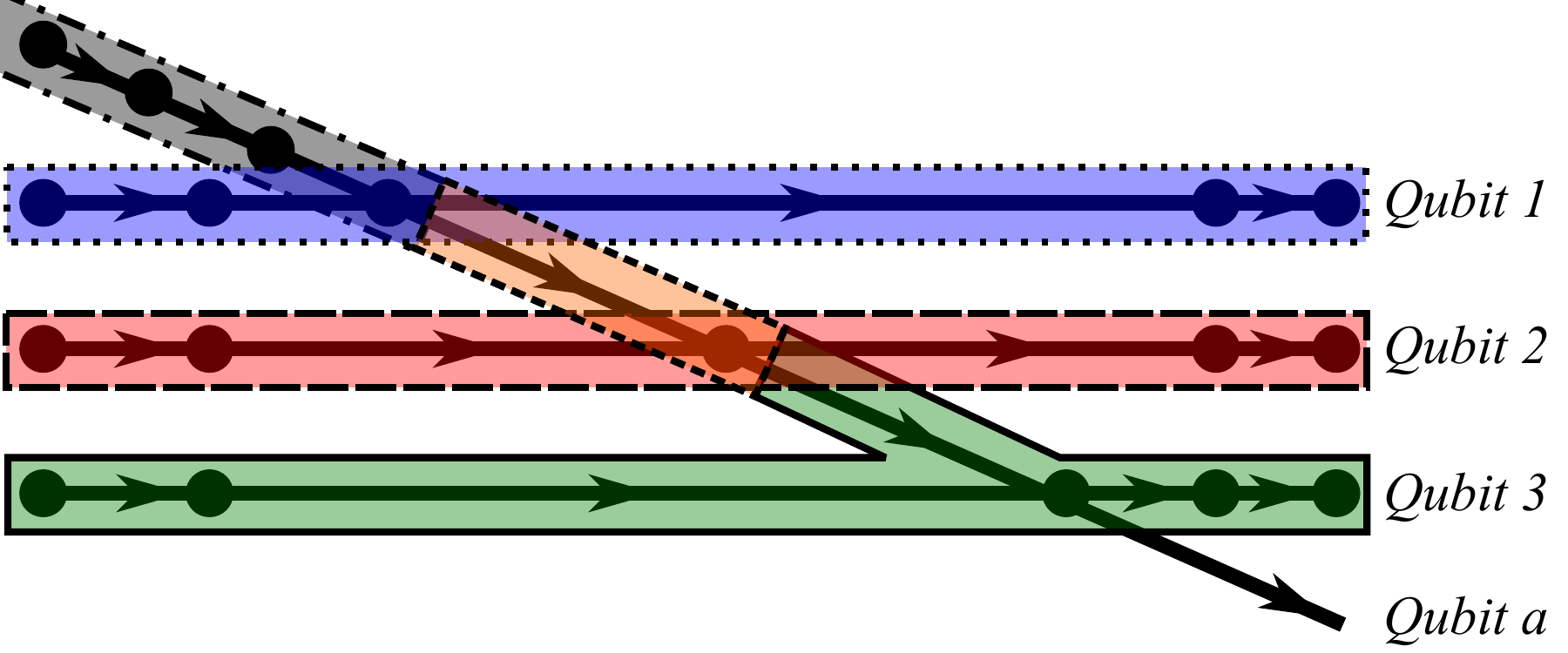}
	\caption{\label{fig:tree} \textbf{a} The Bernstein-Vazirani Algorithm for three bits, with a Hamming weight of three, showing how errors spread in the circuit. Only part of the fault-paths are highlighted to emphasize the tree-pattern formed from fault-paths.  \textbf{b} The same circuit represented as a directed-graph with the full fault-path labeled.}
\end{figure}

The Bernstein-Vazirani Algorithm finds the value of an unknown string, $s$, composed of $m$ unknown bits \cite{BValgo}. It requires the oracle operation $U_{BV}(s)$ that changes the ouput qubit state $y$ based on the data qubits $x$ and the function $f_s(x)$:

\begin{eqnarray*}\label{eq:bv_oracle}
	f_s(x) = \vec{x} \cdot \vec{s} &=& \left(   x_0s_0 + x_1s_1 + \dots + x_{n-1}s_{n-1}  \right)\mod 2 \\
	U_{BV}(s) \ket{x} \ket{y} &=& \ket{x} \ket{y\oplus f_s(x)}.
\end{eqnarray*}

Like all oracle based algorithms, the construction of the oracle is not specified.  We choose the simplest oracle that consists of CNOTs between data qubits where the value of $s$ is 1 and the output qubit. The number of gates depends on the Hamming weight of $s$ and, to determine worst case probabilities, we assume that $s$ has maximum Hamming weight.  

Classically, one sends in data strings with a single bit flipped and determines $s$ in $m$ steps.  Quantum mechanically, by using Hadamard transformations and a Pauli $Z$, one can obtain $s$ in a single oracle call. For this procedure, success is having no bit flips on the data qubits. The output qubit is free to have any error.

Each of the data qubits is measured in the $Z$ basis, implying that only $X$/$Y$ errors are malignant. $\fp_Z(q)$ for each qubit is found using Construction \ref{con:fault-path}. Fig. \ref{fig:faultpath_example} shows the fault-path branching due to the multi-qubit gate. By mapping the overlap between all of the fault-paths, a tree-structure emerges. To emphasize the tree-structure in Fig. \ref{fig:tree}a, some fault-points were deliberately left unhighlighted. This tree-structure meets the main assumption that none of the branches cross each other. To find the success rate for this 3-qubit circuit, each highlighted portion is analyzed separately. Construction \ref{con:qubit_success} gives the probability error state of the Q123 region, which represents the fault-points that affect all three data qubits. By tensoring this state with a unit vector, the first CNOT error matrix can be applied to this state to produce a 16-dimension vector. This larger vector can be divided into four distinct cases: no errors ($\bar{\epsilon}$), error occurring on control branch ($\epsilon_{c}$), error occurring on target branch ($\epsilon_{t}$), and error occurring on both branches ($\epsilon_{ct}$):

\begin{equation}\label{eq:black_vector}
	\begin{bmatrix}
		\bar{\epsilon}\\
		\epsilon_{c} \\
		\epsilon_{t} \\
		\epsilon_{ct}
	\end{bmatrix}_{Q123}
\end{equation}

\noindent After the overlap, each branch is calculated recursively. Since the control branch only contains one fault-path, the probability of no error, $\bar{\epsilon}_{Q1}$, can be found using Construction \ref{con:fault-path}. The target branch contains two fault-paths which have a second overlap region and two additional branches. Similar to the Q123 region, a four-case vector can be found for the Q23 region:

\begin{equation}\label{eq:orange_vector}
	\begin{bmatrix}
		\bar{\epsilon}\\
		\epsilon_{c} \\
		\epsilon_{t} \\
		\epsilon_{ct}
	\end{bmatrix}_{Q23}
\end{equation}

\noindent Similar to before, after the Q23 overlap, the control and target branches have one fault-path each. The probability of no error, $\bar{\epsilon}_{Q2}$ and $\bar{\epsilon}_{Q3}$ respectively, is found using Construction \ref{con:fault-path}. All of these error rates can be combined using Eq. \ref{eq:overlap_matrix} to find the success rate. By dividing the circuit into parts depending on the nodes, the matrices do not change size regardless of the number of qubits.

\begin{equation*}
	\begin{bmatrix}
		\bar{\epsilon} & \epsilon_{c} & \epsilon_{t} & \epsilon_{ct} \\
		\epsilon_{c} & \bar{\epsilon} & \epsilon_{ct} & \epsilon_{t} \\
		\epsilon_{t} & \epsilon_{ct} & \bar{\epsilon} & \epsilon_{c} \\
		\epsilon_{ct} & \epsilon_{t} & \epsilon_{c} & \bar{\epsilon}
	\end{bmatrix}_{Q2,Q3}
	\begin{bmatrix}
		\bar{\epsilon}\\
		\epsilon_{c} \\
		\epsilon_{t} \\
		\epsilon_{ct}
	\end{bmatrix}_{Q23}
	=
	\begin{bmatrix}
		\bar{\epsilon}\\
		~ \\
		~ \\
		\epsilon
	\end{bmatrix}_{Q2,Q3,Q23}
\end{equation*}

\begin{equation}\label{eq:overlap_matrix}
	\begin{bmatrix}
		\bar{\epsilon} & \epsilon_{c} & \epsilon_{t} & \epsilon_{ct} \\
		\epsilon_{c} & \bar{\epsilon} & \epsilon_{ct} & \epsilon_{t} \\
		\epsilon_{t} & \epsilon_{ct} & \bar{\epsilon} & \epsilon_{c} \\
		\epsilon_{ct} & \epsilon_{t} & \epsilon_{c} & \bar{\epsilon}
	\end{bmatrix}_{(Q1),(Q2,Q3,Q23)}
	\begin{bmatrix}
		\bar{\epsilon}\\
		\epsilon_{c} \\
		\epsilon_{t} \\
		\epsilon_{ct}
	\end{bmatrix}_{Q123}
	=
	\begin{bmatrix}
		Success\\
		~ \\
		~ \\
		Error
	\end{bmatrix}
\end{equation}

As with the lowerbound method, various other sub-sets of the Pauli Channel can be found by exchanging $\epsilon$s and $\bar{\epsilon}$s. For example, consider the scenario that the first and third qubit have no error, but the second qubit does have error. To solve for this probability only a minor exchanging of the error rates for the second qubit, $\bar{\epsilon}_{r}$ and $\epsilon_r$, are necessary:

\begin{equation*}
	\begin{bmatrix}
		\bar{\epsilon} & \epsilon_{c} & \epsilon_{t} & \epsilon_{ct} \\
		\epsilon_{c} & \bar{\epsilon} & \epsilon_{ct} & \epsilon_{t} \\
		\epsilon_{t} & \epsilon_{ct} & \bar{\epsilon} & \epsilon_{c} \\
		\epsilon_{ct} & \epsilon_{t} & \epsilon_{c} & \bar{\epsilon}
	\end{bmatrix}_{Q2,Q3}
	\rightarrow
	\begin{bmatrix}
		\epsilon_{c} & \bar{\epsilon} & \epsilon_{ct} & \epsilon_{t} \\
		\bar{\epsilon} & \epsilon_{c} & \epsilon_{t} & \epsilon_{ct} \\
		\epsilon_{ct} & \epsilon_{t} & \epsilon_{c} & \bar{\epsilon} \\
		\epsilon_{t} & \epsilon_{ct} & \bar{\epsilon} & \epsilon_{c}
	\end{bmatrix}_{Q2,Q3}
\end{equation*}

\subsection{\label{sec:QECC}Steane-Shor QECC}

The FPT method was previously used to evaluate syndrome extraction methods for the Steane code on a model ion trap architecture \cite{TomitaPRA2013}. Here we describe the details of the process for a specific syndrome extraction method assuming a quantum machine without geometry, i.e. two-qubit gates are possible between any qubits. The presented FPT method for quantum error correction is an extension and generalization of the previous method described in Ref. \cite{Tomita} and used in Ref. \cite{TomitaPRA2013}. 

For distance-3 codes, all single qubits errors can be decoded. For the Steane Code, $X$ and $Z$ errors are decoded independently, allowing for some two-qubit errors to be fixed. This means the success rate is the probability of all data qubits having less than two errors of the same tpye on two different qubits after the correction is applied. Unlike before, this rate allows multiple correlated output errors, which renders the previous methods inefficient. To reduce the size of the circuit, every syndrome is assumed to be independent, which means they can be analyzed separately. The syndrome is divided into three sub-groups: detectable fault-paths, $S_d$, undetectable fault-paths, $S_u$, and ancilla fault-paths, $S_a$. Detectable fault-paths are sub-groups of data fault-paths where errors will affect the ancilla measurement. In contrast, undetectable fault-paths are those fault-points were the errors will not affect the ancilla measurement. Finally, ancilla fault-paths are the complete fault-paths from ancilla qubits. For our FPT method, we assume these three categories share no fault-points in common. This implies that a single error in any of the three sub-groups will result in a single data-qubit error. Since each FPT calculation is dependent on the individual gate errors, the fault path only produces pseudothreshold curve, not a real threshold point. To find the real threshold, the circuit is encoded to a $k$-level and the error matrices are recursively modified to reflect the $k-1$ error rate. The method is outlined in Construction \ref{con:QECC_success}.

\begin{construct}{\label{con:QECC_success} Probability of QECC successful}{The code style, the error correcting style, and the current level, $k$.}{The probability of error at level $k+1$, $\varepsilon$.}
	\item If the matrix dictionary is not populated at level $k$, populate it by calculating rates for all gates with the level $k-1$ QECC circuit.
	\item Based on the code style and the error correcting style, make the circuit, $\mathbb{C}$.
	\item \textbf{for error\_type in [X, Z]}
	\item \hspace{10pt} Find all fault-paths for data qubits $\rightarrow$ D.
	\item \hspace{10pt} Find all fault-paths for ancilla qubits $\rightarrow$ A.
	\item \hspace{10pt} \textbf{for path in D do}
	\item \hspace{10pt} \hspace{10pt} Separate path into $S_d$ and $S_u$
	\item \hspace{10pt} \textbf{end for}
	\item \hspace{10pt} \textbf{for path in A do}
	\item \hspace{10pt} \hspace{10pt} Separate path into $S_a$ and benign fault-points
	\item \hspace{10pt} \textbf{end for}
	\item \hspace{10pt} Find $\varepsilon_{d}$, $\varepsilon_{u}$, and $\varepsilon_{a}$
	\item \hspace{10pt} $\overline{\varepsilon_{error\_type}}$ = $\overline{\varepsilon_{d}}\overline{\varepsilon_{u}}\overline{\varepsilon_{a}} + \varepsilon_{d}\overline{\varepsilon_{u}}\overline{\varepsilon_{a}} + \overline{\varepsilon_{d}}\varepsilon_{u}\overline{\varepsilon_{a}} + \overline{\varepsilon_{d}}\overline{\varepsilon_{u}}\varepsilon_{a}$
	\item \textbf{end for}
	\item $(1-\varepsilon) = (1-\varepsilon_X) (1-\varepsilon_Z)$
\end{construct}

\begin{figure}[tb]
	\includegraphics{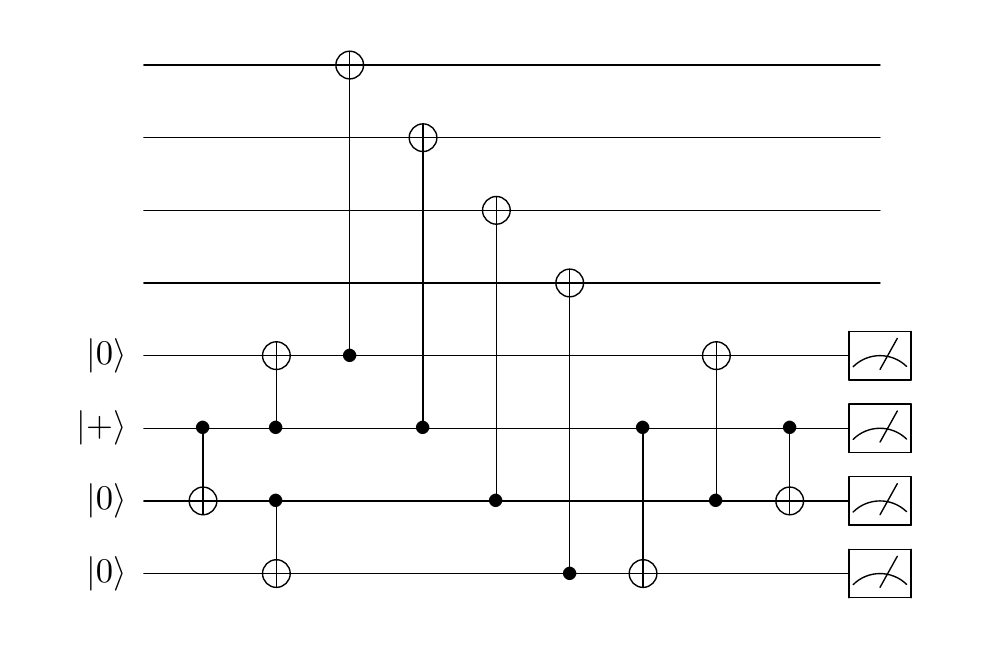}
	\caption{\label{fig:steane_shor} A single syndrome measurement for the Steane-Shor QEC with DiVincenzo decoding. The method generates a undirected cycle in the circuit diagram precludng the use of our methods for tree-like circuits.}
\end{figure}

The exact procedure to find $\varepsilon_{d}$, $\varepsilon_{u}$, and $\varepsilon_{a}$ varies with each QECC.   Here we describe how it is applied to Steane QECC with Shor ancilla and the decoding scheme proposed by Divincenzo and Aliferis to account for the overlap between $S_d$ and $S_a$ in each syndrome  \cite{Steane}. An example sysdrome measurement circuit is shown in Fig. \ref{fig:steane_shor}.
This QECC measures each syndrome ($X$ and $Z$) three times, and employs a majority vote to ensure accurate corrections. Since each syndrome is independent, calculations can be reduced by assuming $\varepsilon_{d1} = \varepsilon_{d2} = \varepsilon_{d3}$. For each syndrome, the fault-paths for the DiVincenzo and Aliferis correction are found first. Based on the probability that an error will spread to both the ancilla measurements and the data measurements, additional gates are added to the data qubits to represent the probability of a correction occurring. For the case of the Steane-Shor QECC, the detectable and ancilla groups have a number of shared fault-points; therefore, the overlap between these groups is treated as a forth group, $S_b$. The data qubit fault-paths are divided among undetectable and detectable while the remaining ancilla fault-paths remain intact. Construction \ref{con:gen_success} is used to find $\varepsilon_{d}$, $\varepsilon_{u}$, $\varepsilon_{b}$, and $\varepsilon_{a}$. For this particular QECC, a single error in any of the four categories will render the entire syndrome faulty. Using the probability of a single $X$ and $Z$ syndrome measuring fault, the probability of the three syndromes giving the right correction is easy to calculate. 

In general, this method is accurate when there is very little or no overlap between $S_d$ and $S_a$. In addition, many QECCs require decoding schemes to reduce the number of relevant qubits and account for any classical computations. Without these decoding schemes, the number of possible outcomes quickly renders the FPT method ill-suited. In general, the FPT method cannot simultaneously calculate multiple parts of the Pauli channel. To find the full Pauli channel exactly requires $G$ $4^m \times 4^m$ matrices where $G$ is the number of gates and $m$ is the number of data and ancilla qubits. These matrices would act on a size $4^m$ probability error state vector. Any correction steps would also need to be represented as $4^m \times 4^m$ matrices, as no classical corrections can be applied in this context.

\section{\label{sec:results}Results}

All matrix and vector math is done using the NumPy python package \cite{NumPy}.

\subsection{\label{sec:BV_results}Bernstein-Vazirani Algorithm}

\begin{figure}[tb]
	\begin{center}
		(a) \includegraphics[width = 0.7\textwidth]{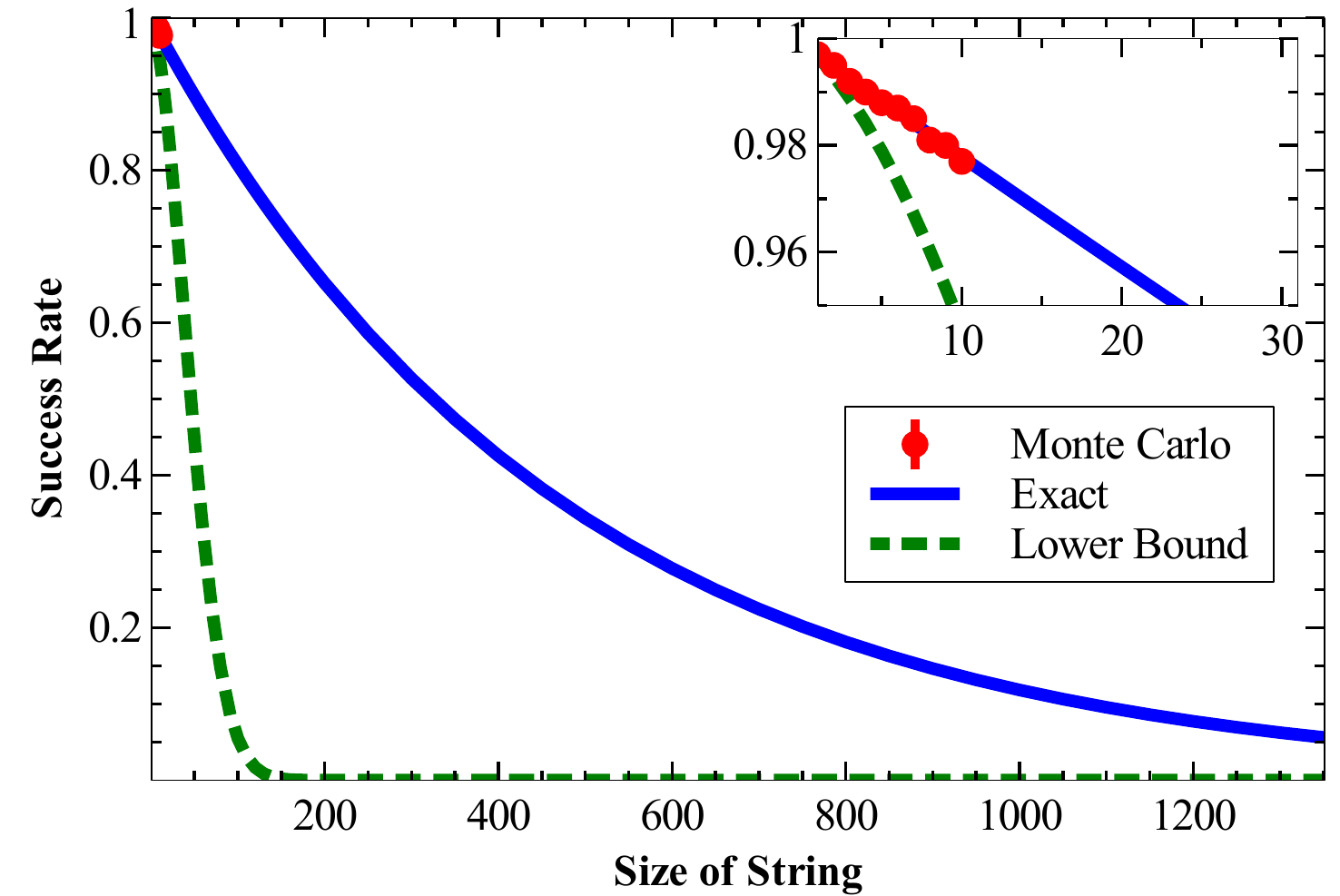} 
		
		(b) \includegraphics[width = 0.7\textwidth]{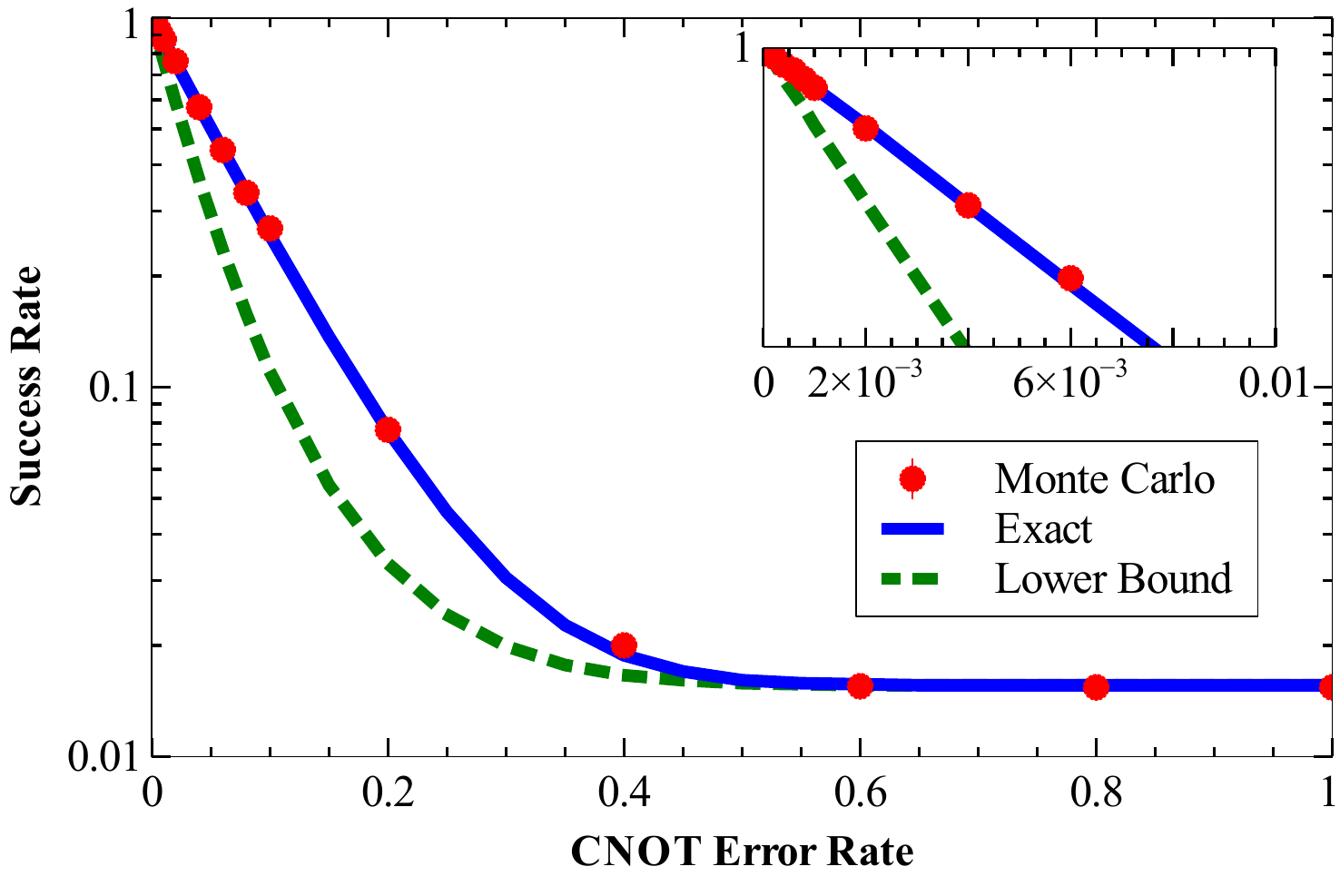}
	\end{center}
	\caption{\label{fig:accuracy}\textbf{a} Comparison of exact and approximate FPT methods to Monte Carlo for the Bernstein-Vazirani algorithm with CNOT error rate = $1.0\times 10^{-3}$ and Hamming weight equal to the string size. \textbf{b}  Here we vary the error rate for a string size and Hamming weight equal to 6.}
\end{figure}

For testing purposes, we choose to model error as Markovian-depolarizing noise. Depolarizing noise represents the error rate of all gates as $\epsilon$. Since single-qubit gates have three types of error ($X$, $Y$, and $Z$), each type of error has an equal chance of occurring ($\frac{\epsilon}{3}$). For two-qubit gates, this fraction changes to $\frac{\epsilon}{15}$ to represent the additional types of error ($XX$, $YZ$, etc.).

When comparing the fault-path method to Monte Carlo simulations, there are two main parameters: accuracy of the success rate and computation speed. We tested both of these parameters against two circuit variables: the gate error rate, $\epsilon$, and the size of the unknown string, $s$. The Monte Carlo results consisted of many trials. Each ($\epsilon$,$s$) combination was simulated $\frac{10 \cdot s}{\epsilon}$ times with a minimum of 100,000, and each combination is an average of least three trials. As seen in Fig. \ref{fig:accuracy}, the success rate behavior is reasonable since it decreases for higher error rates and increases for smaller circuit sizes.

The exact FPT method accurately predicts all Monte Carlo results, both when the size of the string and the error rate are varied, Fig. \ref{fig:accuracy}.  In contrast, the lowerbound FPT method has regions of ($\epsilon$,$s$) that appear more accurate. As the string size increases, the lowerbound method loses accuracy at an exponential rate. Comparatively, at error rates less than 0.002 and higher than 0.4, the percent error is less than 5\%, while the region in between has percent error as high as 60\%. In general, the lowerbound method reasonable predict the correct success rate with a percent error less than 5\% at $s\cdot \epsilon <$ 0.01.

Both the exact FPT method and the lowerbound FPT method consistently take less time as expected from an analytical method. As the fault-path method for tree-circuits is fully independent of error rate, the timing does not change based on error rate, unlike Monte Carlo methods.

\subsection{\label{sec:QECC_results}Steane-Shor QECC}

\begin{figure}[tb]
	\includegraphics[width = 0.95\textwidth]{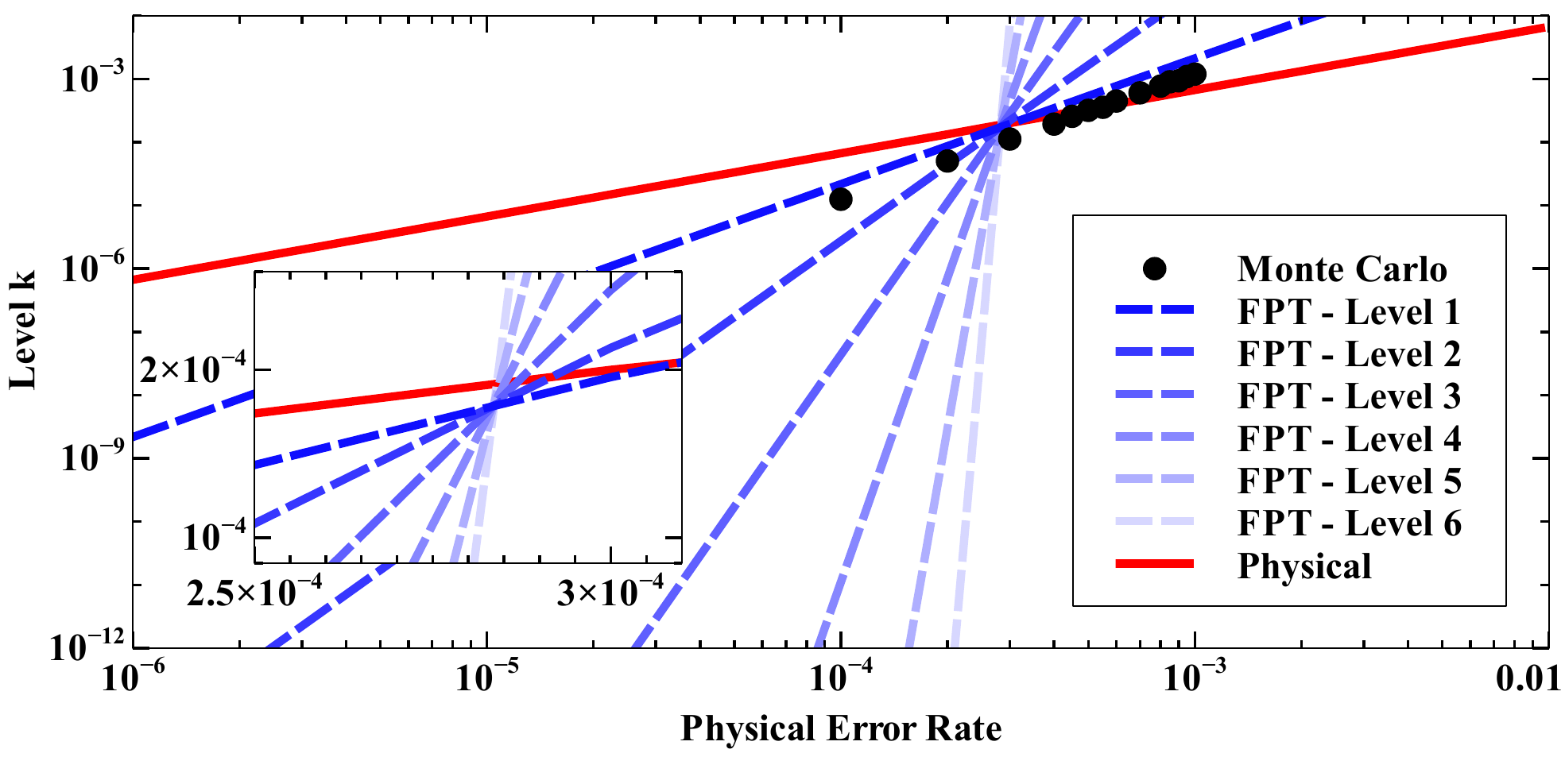}
	\caption{\label{fig:steane} The threshold curves based on two methods: the FPT and Monte Carlo simulations for a EC circuit. The FPT shows the threshold curve for different levels of encoding to find the threshold. MC results were found at level one. The AGP result represents the predicted threshold at an infinite level.}
\end{figure}

A key figure for any error-correcting code is where the logical error rate is less than the physical error rate. This first error threshold is called the pseudothreshold.  The threshold is defined for a code family and is the error which below one can achieve arbitrary low failure probability by increasing the code distance.  Fig. \ref{fig:steane} compares the FPT method to Monte-Carlo. We expect Monte-Carlo to give exact results but also it requires more statistics as error rates are reduced \cite{new}. Here we use it to benchmark the pseudothreshold for an isolated error correction implementation.  We see that the FPT method yields similar results.

Using the fault-path tracer method, the threshold curve was found for the first five levels of encoding, Fig. \ref{fig:steane}. The Steane-Shor circuit does not follow the binary-tree pattern; therefore, the FPT method only produces a lower bound on the threshold. It estimates the pseudothreshold at $3.25\times 10^{-4}$ which is lower than the Monte Carlo simulations. Since the difference between these two curves is a second degree polynomial, this emphasizes how the tracing method misses some errors that cancel. Under the assumptions that logical measurement and preparation operations have failure rates as if they were transversal, a level $k$ circuit can be analyzed in terms of $k-1$-level error rates. Our method estimates the real threshold at $1.91\times 10^{-4}$. Here we examine a circuit of I followed by error correction. 

The method of Aliferis, Gottesman, and Preskill (AGP) based on fault-paths and malignant pair counting produces a conservative estimate of the threshold. We implemented the AGP method using code from Andrew Cross \cite{Cross}. We were able to predict a memory threshold assuming error correction, an identity gate, and then error correction.  We found a threshold of $5.91 \times 10^{-5}$.  We expect that the real Pauli error threshold lies above our estimate and this estimate.

\section{Conclusions}

The analytic methods based on fault paths can be used to accurately assess the integrated performance of quantum devices. We have shown the utility of fault paths for understanding the failure of simple algorithms and error correcting codes. Although the method is limited to circuits which are not universal with relatively simple structures, the method is scalable to many qubits. We expect that testing the performance of faulty quantum computers on easy problems will be an important method for showing that errors between gates are sufficiently independent for error correction to work.   

The work also suggests that a tensor network approach could be applied to calculate the error of the circuits \cite{OrusAnnPhys2014}. Tensor networks are typically used to describe quantum states and to calculate their properties. In this case the tensor network describes the error states and  sampling different error output configurations would correspond to changing output error vectors. We expect similarities with the graphical methods for stabilizer circuits \cite{Faster_than_CHP}. Tensor network contraction also naturally allows for partial parallelization of algorithms and this may lead to  faster algorithms for more accurate estimation of error correcting circuit thresholds.

\begin{acknowledgements}
	We would like to thank Andrew Cross, Ryan Sheehan, Alonzo Hernandez, and Silas Fradley for useful discussions.  This project was supported by Office of the Director of National Intelligence (ODNI) - Intelligence Advanced Research Project Activity (IARPA) through Army Research Office Grant No. W911NF-10-1-0231 and Department of Interior Contract No. D11PC20167. All statements of fact, opinion or conclusions contained herein are those of the authors and should not be construed as representing the official views or policies of IARPA, the ODNI, or the U.S. Government.
\end{acknowledgements}


\begin{thebibliography}{10}
	\providecommand{\url}[1]{{#1}}
	\providecommand{\urlprefix}{URL }
	\expandafter\ifx\csname urlstyle\endcsname\relax
	\providecommand{\doi}[1]{DOI \discretionary{}{}{}#1}\else
	\providecommand{\doi}{DOI \discretionary{}{}{}\begingroup
		\urlstyle{rm}\Url}\fi
	
	
	\bibitem{Pauli_twirling_Chuang}
	I.L. Chuang, M.A. Nielsen, Prescription for experimental determination of the dynamics of a quantum black box, J. Mod. Opt. \textbf{44}, 2455 (1997)
	
	\bibitem{RB}
	Emerson, J., Alicki, R., \.{Z}yczkowski, K.: Scalable noise estimation with random unitary operators, J. Opt. B \textbf{7}, S347-S352 (2005)
	
	\bibitem{Emerson07}
	Emerson, J., Silva, M., Moussa, O., Ryan, C., Laforest, M., Baugh, J., Cory, D.G.,
	Laflamme,R.: Symmetrized characterization of noisy quantum processes, Science \textbf{317}, 1893 (2007)
	
	\bibitem{RB3}
	L\'{e}vi, B., L\'{o}pez, C. C., Emerson, J., Cory, D. G.: Efficient error characterization in quantum information processing, Phys. Rev. A \textbf{75}, 022314 (2007)
	
	\bibitem{RB4}
	Dankert, C., Cleve, R., Emerson, J., Livine, E.: Exact and approximate unitary 2-designs: constructions and applications, Phys. Rev. A \textbf{80}, 012304 (2009)
	
	\bibitem{Rand_Knill}
	Knill, E., Leibfried, D., Reichle, R., Britton, J., Blakestad, R.B., Jost, J.D.,
	Langer, C., Ozeri, R., Seidelin, S., Wineland, D.J.: Randomized benchmarking of quantum gates, Phys. Rev. A \textbf{77},
	012307 (2008)
	
	\bibitem{Flammia_tomo}
	Flammia, S.T., Liu, Y.K.: Direct fidelity estimation from few Pauli measurements, Phys. Rev. Lett. \textbf{106}, 230501 (2011)
	
	\bibitem{Flammia_tomo2}
	Flammia, S. T., Gross, D., Liu, Y.K., Eisert, J.: Quantum tomography via compressed sensing: error bounds, sample
	complexity and effcient estimators, New J. Phys. \textbf{14}, 095022 (2012)
	
	\bibitem{Poulin_tomo}
	da~Silva, M.P., Landon-Cardinal, O., Poulin, D.:Practical characterization of quantum devices without tomography, Phys. Rev. Lett. \textbf{107},
	210404 (2011)
	
	\bibitem{BlumeKohout2013}
	Blume-Kohout, R., Gamble, J.K., Nielsen, E., Mizrahi, J., Sterk, J.D., Maunz, P.: Robust, self-consistent, closed-form tomography of quantum logic gates on a trapped ion qubit,
	arXiv:1310.4492  (2013)
	
	\bibitem{MerkelPRA2013}
	Merkel, S.T., Gambetta, J.M., Smolin, J.A.,  Poletto, S., C\'orcoles, A.D.,
	Johnson, B.R., Ryan, C.A., Steffen, M.: Self-consistent quantum process tomography, Phys. Rev. A \textbf{87}, 062119 (2013)
	
	\bibitem{WallmanNJP2015}
	Wallman, J., Granade, C., Harper, R., Flammia, S.T.: Estimating the coherence of noise, New J. Phys. \textbf{17}(11),
	113020 (2015)
	
	\bibitem{qdyn_comp_sens}
	Shabani, A., Kosut, R.L., Mohseni, M., Rabitz, H., Broome, M.A., Almeida, M.P.,  Fedrizzi, A., White, A.G.: Efficient measurement of quantum dynamics via compressive sensing, Phys. Rev. Lett. \textbf{106}, 100401 (2011)
	
	\bibitem{RB5}
	Magesan, E., Gambetta, J.M., Emerson, J.: Robust randomized benchmarking of quantum processes, Phys. Rev. Lett. \textbf{106} 180504 (2011)
	
	\bibitem{RB6}
	Magesan, E., Gambetta, J.M., Emerson, J.: characterizing quantum gates via randomized benchmarking, Phys. Rev. A \textbf{85} 042311 (2012)
	
	\bibitem{HarrowPRL2009}
	Harrow, A.W., Hassidim, A., Lloyd, S.: Quantum algorithm for linear systems of equations, Phys. Rev. Lett. \textbf{103}, 150502
	(2009)
	
	\bibitem{CladerPRL2013}
	Clader, B.D., Jacobs, B.C., Sprouse, C.R.: Preconditioned quantum linear system algorithm, Phys. Rev. Lett. \textbf{110}, 250504
	(2013)
	
	\bibitem{CHP}
	Aaronson, S., Gottesman, D.: Improved simulation of stabilizer circuits, Phys. Rev. A \textbf{70}, 052328 (2004)
	
	\bibitem{Faster_than_CHP}
	Anders, S., Briegel, H.J.: Fast simulation of stabilizer circuits using a graph-state representation, Phys. Rev. A \textbf{73}, 022334 (2006)
	
	\bibitem{Cory}
	Magesan, E., Puzzuoli, D., Granade, C.E., Cory, D.G.: Modeling quantum noise for efficient testing of fault-tolerant circuits, Phys. Rev. A \textbf{87},
	012324 (2013)
	
	\bibitem{PRA_us}
	Guti\'errez, M.,  Svec, L.,  Vargo, A.,  Brown, K.R.: Approximation of realistic errors by Clifford channels and Pauli measurements, Phys. Rev. A \textbf{87}, 030302
	(2013)
	
	\bibitem{GutierrezPRA2015}
	Guti\'errez, M., Brown, K.R.: Comparison of a quantum error-correction threshold for exact and approximate errors, Phys. Rev. A \textbf{91}, 022335 (2015)
	
	\bibitem{AGP_method}
	Aliferis, P.,  Gottesman, D.,  Preskill, J.: Quantum accuracy threshold for concatenated distance-3 codes, Quant. Inf. Comput. \textbf{6}, 97
	(2006)
	
	\bibitem{BValgo}
	Bernstein, E., Vazirani, U.: Quantum complexity theory, in Proceedings of the Twenty-fifth Annual ACM
	Symposium on Theory of Computing (ACM, New York, NY, USA, 1993), STOC '93,
	pp. 11--20
	
	\bibitem{Steane_QEC}
	Steane, A.M., Error correcting codes in quantum theory, Phys. Rev. Lett. \textbf{77}, 793 (1996)
	
	\bibitem{TomitaPRA2013}
	Tomita, Y.,  Guti\'errez, M.,  Kabytayev, C., Brown, K.R.,  Hutsel, M.R., Morris, A.P., Stevens, K.E., Mohler, G.: Comparison of ancilla preparation and measurement procedures for the Steane [[7,1,3]] code on a model ion-trap quantum computer, Phys. Rev. A \textbf{88}, 042336 (2013)
	
	\bibitem{MikeandIke}
	Nielsen, M.A.,  Chuang, I.L.: Quantum computation and quantum information,
	10th edn. (Cambridge University Press, 2010)
	
	\bibitem{Tomita}
	Tomita, Y.: Numerical and analytical studies of quantum error correction.
	\newblock Ph.D. thesis, Georgia Institute of Technology (2014)
	
	\bibitem{Steane}
	DiVincenzo, D.P., Aliferis, P.: Effective fault-tolerant quantum computation with slow measurements, Phys. Rev. Lett. \textbf{98}, 220501 (2007)
	
	\bibitem{Cross}
	Aliferis, P., Cross, A.W.: Subsystem fault tolerance with the Bacon-Shor code, Phys. Rev. Lett. \textbf{98}, 220502 (2007)
	
	\bibitem{NumPy}
	Walt, S. v. d., Colbert, S. C., Varoquaux, G.: The NumPy array: a structure for efficient numerical computation, CiSE \textbf{13}, 22-30 (2011)
	
	\bibitem{OrusAnnPhys2014}
	Or\'us, R.: A practical introduction to tensor networks: Matrix product states and projected entangled pair states, Ann. Phys. \textbf{349}, 117-158 (2014)
	
	\bibitem{new}
	Bravyi, S., Vargo, A.: Simulation of rare events in quantum error correction, Phys. Rev. \textbf{88}, 062308 (2013)
	
\end{thebibliography}
\end{document}